\documentclass[apjl]{emulateapj}
\usepackage{color}
\usepackage{rotating}
\usepackage{subfigure}
\usepackage{natbib}
\shorttitle{SN~2010gx host}
\shortauthors{Chen et al.}

\begin{document}

\title{The host galaxy of the super-luminous SN~2010gx and limits on explosive $^{56}$Ni production}

\author{
Ting-Wan Chen\altaffilmark{1}; 
Stephen J. Smartt\altaffilmark{1}; 
Fabio Bresolin \altaffilmark{2}; 
Andrea Pastorello\altaffilmark{3}; 
Rolf-Peter Kudritzki \altaffilmark{2}; 
Rubina Kotak \altaffilmark{1}; 
Matt McCrum \altaffilmark{1}; 
Morgan Fraser \altaffilmark{1};
Stefano Valenti \altaffilmark{4,5}
}

\altaffiltext{1}{Astrophysics Research Centre, School of Mathematics and Physics, Queen's University Belfast, Belfast BT7 1NN, UK}
\altaffiltext{2}{Institute for Astronomy, 2680 Woodlawn Drive, Honolulu, HI 96822, USA}
\altaffiltext{3}{INAF-Osservatorio Astronomico di Padova, vicolo dell'Osservatorio 5, 35122 Padova, Italy}
\altaffiltext{4}{Las Cumbres Observatory Global Telescope Network, Inc. Santa Barbara, CA 93117, USA}
\altaffiltext{5}{Department of Physics, University of California Santa Barbara, Santa Barbara, CA 93106-9530, USA}

\slugcomment{Accepted for publication in Astrophysical Journal Letters, 2012 Dec 27}

\begin{abstract}
Super-luminous supernovae have a tendency to occur in faint host galaxies which are 
likely to have low mass and low metallicity. While these extremely luminous explosions 
have been observed from $z=0.1$ to $1.55$, the closest explosions allow more detailed 
investigations of their host galaxies. We present a detailed analysis of the host galaxy 
of SN~2010gx ($z=0.23$), one of the best studied super-luminous type Ic supernovae. 
The host is a dwarf galaxy ($M_{g}=-17.42\pm0.17$) with a high specific star formation 
rate. It has a remarkably low metallicity of $12+\log {\rm(O/H)} = 7.5\pm0.1$ dex as 
determined from the detection of the [OIII] $\lambda4363$ line. This is the first reliable 
metallicity determination of a super-luminous stripped-envelope supernova host.
We collected deep multi-epoch imaging with Gemini + GMOS between $240-560$ days 
after explosion to search for any sign of radioactive $^{56}$Ni, which might provide
further insights on the explosion mechanism and the progenitor's nature. We reach 
$griz$ magnitudes of $m_{\rm AB}\sim26$, but do not detect SN~2010gx at these 
epochs. The limit implies that any $^{56}$Ni production was similar to or below 
that of SN~1998bw (a luminous type Ic SN that produced around 0.4\,$M_{\odot}$ of
$^{56}$Ni). The low volumetric rates of these supernovae ($\sim10^{-4}$ of the 
core-collapse population) could be qualitatively matched if the explosion mechanism 
requires a combination of low-metallicity (below $0.2\,Z_{\odot}$), high progenitor mass 
($>60\,M_{\odot}$)  and high rotation rate (fastest 10\% of rotators). 
\end{abstract}


\section{Introduction}
Novel wide-field survey projects, such as the Panoramic Survey Telescope And Rapid 
Response System (Pan-STARRS-1) \citep{2010SPIE.7733E..12K} and the Palomar 
Transient Factory (PTF) \citep{2009PASP..121.1334R}, have carried out searches for 
supernovae without an inherent galaxy bias. \citet{2011Natur.474..487Q} identified 
a class of hydrogen-poor ``super-luminous supernovae (SLSNe)'' with typical 
$g-$band absolute magnitudes of around $-21.5$ , which is 10 to 100 times brighter 
than normal core-collapse SNe. Their intrinsic brightness has allowed them to be 
discovered at high redshift in the Pan-STARRS-1 survey between $z=1-1.55$ 
\citep{2011ApJ...743..114C, 2012ApJ...755L..29B}. \citet{2010ApJ...724L..16P} studied 
one of the closest examples, SN~2010gx ($z=0.23$), in detail showing it to transition to 
a normal type Ic SN at 40 days after peak, while multi-colour photometry rules out 
$^{56}$Ni as powering the peak magnitude. Two physical mechanisms for SN~2010gx 
(and the supernovae like it from \citeauthor{2011Natur.474..487Q,2011ApJ...743..114C}) 
have been proposed. One is that the type Ic SN is boosted in luminosity by energy deposition 
from magnetar spin down \citep{2010ApJ...717..245K}. The second is the shock breakout 
through a dense wind \citep{2011ApJ...729L...6C}, or shock-interaction of the SN ejecta 
with massive C-O shells \citep{2010arXiv1009.4353B}. 

\citet{2011ApJ...727...15N} showed that most SLSNe tend to occur in star-forming dwarf 
galaxies, which have high specific star formation rates. Their intrinsic rate is 
$10^{-8}$\,Mpc$^{-3}$\,year$^{-1}$; or $10^{-4}$ of the normal core-collapse SN rate
 \citep{2011Natur.474..487Q,2012Sci...337..927G}. An important question is whether or 
not the apparent preference they have for dwarf host galaxies is related to requiring low 
metallicity progenitor stars. As low metallicity affects stellar structure and evolution 
through changes in stellar winds, opacity and rotation, quantitative measurements are 
required. In this letter we study the host galaxy of one of the nearest SLSNe, SN~2010gx, 
and search for evidence of late emission from radioactive $^{56}$Ni decay.

\section{Observations}
\subsection{Photometry}
We used Gemini South and North Observatories with Gemini Multi-Object Spectrograph (GMOS) 
to collect $griz$ photometry between $240-560$ days after the maximum of SN~2010gx (Table\,1). 
To check for the presence of residual supernova flux at $240-370$ days after peak, we aligned and 
rescaled all images to the same pixel grid using {\sc IRAF}/{\sc geomap} and {\sc geotran} packages. 
We performed imaging subtraction using the High Order Transform of PSF ANd Template 
Subtraction ({\sc hotpants}) software 
\footnote{http://www.astro.washington.edu/users/becker/hotpants.html},
to subtract the January 2012 images from the earlier epochs. The host galaxy subtracts off almost 
perfectly and no SN signal is detected, hence we made the assumption that the final image taken 
in January 2012 only contains flux from the host galaxy. In addition, the deep spectra of the host 
do not show any broad features or evidence commonly associated with SNe spectra.

To determine a limiting  magnitude for the SN flux in the three epochs, we added 6 fake stars, 
one at the supernova position and five within a surrounding radius of 0.6 arcmin to each of 
the $240-370$ days images. We ran this procedure multiple times with a range of fake star
magnitudes and repeated image subtraction with {\sc hotpants}. We determined $5\sigma$ and 
$3\sigma$ detection limits by requiring that we could visually detect a fake SN exactly at the 
position of SN~2010gx and that the measured standard deviation of the set of fake stars added 
was 0.2 and 0.3 magnitudes respectively. The resulting upper limits are in Table\,1.

The apparent magnitudes (SDSS AB system) of the host galaxy, 
SDSS J112546.72-084942.0, are $g=23.71\pm0.14$, $r=22.98\pm0.10$, $i=22.90\pm0.13$ 
and $z=22.98\pm0.16$. These were obtained from aperture photometry 
(within {\sc IRAF}/{\sc daophot}) on the best seeing images, using a zero-point calibration 
achieved with 10 SDSS reference stars. As deep, pre-explosion images of the host galaxy were 
not available, we compared our photometry results to both SDSS catalog magnitudes, and our 
own photometry performed on the SDSS images. In the $g-$band, the SDSS catalog Petrosian 
magnitude for the object is $22.8\pm0.5$, while we measure $22.6\pm0.6$ mag in the 
SDSS image. The large uncertainty is due to the marginal detection. Our deep Gemini image 
shows the host resolved from a fainter neighbouring galaxy (which is unresolved from the 
host in SDSS). If we use a large aperture on our Gemini $g-$band image, to include flux from 
the neighbouring galaxy then we obtain $g=23.29\pm0.16$ mag. Within the uncertainties, 
the SDSS and Gemini magnitudes are similar and there is no hint that the $+560$ days Gemini
images are brighter than the SDSS images and contain SN flux. This supports our 
assumption that the January 2012 images are dominated by galaxy flux and contain no 
detectable, residual SN flux.

\subsection{Spectroscopy}

Spectroscopy of SN~2010gx was taken on 5 June 2010
\citep{2010ApJ...724L..16P} with the Gemini South Telescope+GMOS. The B600 grating 
(ID $G5323$; $2\times1800$s) provided coverage in the observer frame of 
$4300-6400$\,\AA\ (henceforth referred to as the ``blue spectrum''). Although this 
spectrum contains flux from SN~2010gx, we fitted a high order polynomial (order 15) 
through the SN continuum and subtracted off this contribution leaving a flat spectrum for which 
emission lines between $3500-5200$\,\AA\ (rest frame) could be measured. We could not 
measure the H$\alpha$ line in this spectrum due to contamination by a strong sky line. 
A second spectrum was obtained on 23 Dec. 2011 with GMOS at the Gemini North Telescope, 
using the R400 grating (ID $G5305$; $3\times1800$s), to cover $5070-8500$\,\AA\ 
(henceforth referred to as the ``red spectrum''). An order-blocking filter ($GG455$) and low 
grating efficiency truncated the wavelength coverage in the blue. Detrending of the data, such 
as bias-subtraction, was established using standard techniques within {\sc IRAF}. For the red 
spectra, individual exposures were nodded along the slit, and we used two-dimensional image 
subtraction to remove the sky background. Wavelength and flux calibrations were achieved 
using daytime CuAr arcs and the spectrophotometric standard Feige 34.

We used $1.5''$ and $1.0''$ slits for red and blue spectra separately, yielding a resolution 
(as measured from narrow night sky lines) of $\sim6$\,\AA\ in the red, and $\sim4.5$\,\AA\ 
in the blue. The combined blue and red spectra provide spectral coverage between 4300 and 
8500\,\AA\ and the overlap region was used to ensure a uniform flux calibration, employing 
the three strong lines [OIII] $\lambda5007$, [OIII] $\lambda4959$ and H$\beta$. We measured 
the line fluxes in both spectra and applied a linear scaling to bring the blue spectrum into 
agreement with the red. The line fluxes of these three lines agreed to within $3\%$ after this 
rescaling.

Line flux measurements were made after fitting a polynomial function to subtract the continuum. 
Fig.\,3 shows the final combined spectrum of the host galaxy. We fitted Gaussian line profiles 
within a custom built {\sc IDL} environment, fixing the full-width-half-maximum (FWHM) of the 
mean of the three strong lines; 6.03\,\AA\ and 4.45\,\AA\ for the red and blue spectra 
respectively (Table\,2). We also normalized the spectrum and determined the line equivalent widths. 
These equivalent widths and rms of the continuum provided uncertainty estimates using the 
expression from \citep[][Table\,2]{2003MNRAS.346..105P}.

\section{Analysis}
\subsection{$^{56}$Ni mass estimation for SN~2010gx}
The fact that SN~2010gx was not detected after the host galaxy image subtraction at epochs 
between $+244$ and $+362$ days allows us to determine an upper limit for the mass of 
$^{56}$Ni ejected. This is somewhat uncertain as it relies on assuming that the gamma ray trapping 
in SN~2010gx behaves like in other Ic SNe in the nebular phase, and the late luminosity is directly 
proportional to the mass of $^{56}$Co present at any epoch. 

We estimated the upper limit for any flux coming from SN~2010gx at the 
effective rest wavelength of the $ri-$filters. These were corrected for Milky
Way extinction only.  
\footnote{Hubble constant of H$_{0}=72$\,km\,s$^{-1}$\,Mpc$^{-1}$, 
$\Omega_{\rm M}=0.3$, $\Omega_{\rm \lambda}=0.7$, gives a luminosity
distance 1115 Mpc for $z=0.23$ \citep{2006PASP..118.1711W}.}
We compared the upper limit of the luminosities of SN~2010gx with the late-time luminosities of several 
type Ic SNe, such as SN~1998bw, SN~2002ap and SN~2007gr 
\citep{2011AJ....141..163C, 2006ApJ...644..400T, 2009A&A...508..371H}.
The flux measured in the observed $ri$ frames of SN~2010gx were
compared with those for the low redshift SNe in Johnson $V$ and
Cousins $R$ as the effective wavelengths roughly match at $z=0.23$ 
(see Fig. 2). 
The most interesting measurement is the $i-$band limit for SN~2010gx flux at 
$+244$ days after the SN explosion, which is similar to or below what we would expect if SN~1998bw were 
at this distance. If we take $0.4\,M_{\odot}$ as the representative $^{56}$Ni mass for SN~1998bw 
\citep{2006ApJ...640..854M}, then the flux limit of SN~2010gx suggests an upper limit of about 
$0.4\,M_{\odot}$ of $^{56}$Ni ejected by SN~2010gx. For a complete comparison with other type Ic SNe, 
we also included SN~2002ap and SN~2007gr. Both these SNe were modelled to indicate that the ejecta contained 
around $0.08$\,M$_{\odot}$ of $^{56}$Ni \citep{2006ApJ...644..400T, 2009A&A...508..371H}.
Our limits are significantly brighter than these two, hence we can simply say that SN~2010gx did not produce 
explosive $^{56}$Ni in larger quantities than the brightest known type Ic SNe such as SN~1998bw.

\subsection{Host galaxy metallicity and other properties}

We carried out an abundance analysis of the host galaxy based on the determination of the electron 
temperature, and detection of the [OIII] $\lambda4363$ auroral line (see \citealt{2009ApJ...700..309B} 
for further details). Firstly, the emission line fluxes were reddening corrected using the Milky Way
law of A$_{V}=3.17\times$\,E($B-V$). We assumed case B recombination which requires an intrinsic line 
ratio of H$\alpha$/H$\beta=2.86$ and H$\gamma$/H$\beta=0.47$ at T$_{e}=10^{4}$\,K
 \citep{1989agna.book.....O}. The observed ratio of H$\alpha$/H$\beta=3.88$ yields an extinction 
coefficient c(H$\beta$)$=0.48$, which we used to calculate reddening-corrected line fluxes. The 
observed line fluxes and errors are listed in Table\,2.

We confidently detected the [OIII] $\lambda4363$ auroral line (S/N$\sim6$) in the GMOS spectrum 
(Fig.\,3), allowing us to estimate the electron temperature and calculate the oxygen abundance
directly. As in \citet{2009ApJ...700..309B}, we used a custom-written {\sc pyraf} script based on 
the {\sc IRAF}/{\sc nebula} package. The direct method of estimating oxygen abundances uses the 
ratio of the intensities of [OIII] $\lambda4959$, $\lambda5007$/[OIII] $\lambda4363$ lines to
determinate the electron temperature (T$_{e}$) in the ionized gas region. We find an oxygen 
abundance of $12+\log {\rm(O/H)}=7.46\pm0.10$ dex, remarkably low even for dwarf galaxies 
(Fig.\,3). Assuming a solar value of $8.69\pm0.05$ dex \citep{2009ARA&A..47..481A}, this implies
that the host galaxy is 1.2 dex below solar abundance or just $0.06\,Z_{\odot}$.
\citet{2011ApJ...730...34S} estimated a metallicity from the strong lined methods, but it is 
well documented that the results from this method can vary by up to 0.7\,dex depending on 
which calibration is used \citep{2009ApJ...700..309B}. Not surprisingly, they get an estimate of 8.16 dex 
from the N2 diagnostic, which appears 0.7 dex systematically too high, when compared with our direct estimate.

The reddening-corrected and $z-$corrected flux of H$\alpha$ is 
$3.49\times10^{-16}$ (erg\,s$^{-1}$\,cm$^{-2}$).
The FWHM of the galaxy was measured at $0.7''$ and the slit was $1.5''$, therefore $>98\%$ of 
the host was completely encompased in the slit, from which we determined the star formation 
rate (SFR) of the host to be $0.41\,M_{\odot}$\,year$^{-1}$ from the calibration of
\citet{1998ARA&A..36..189K}: SFR ($M_{\odot}$\,year$^{-1}$)$=7.9\times$ 
10$^{-42}$ L(H$\alpha$) (ergs\,s$^{-1}$). 
We also estimated the SFR from the [OII] $\lambda3727$ line flux, 
considering slit coverage of the host in the blue arm is $90\%$, 
which gives $0.40\pm0.10\,M_{\odot}$\,year$^{-1}$ \citep{1998ARA&A..36..189K}
and is consistent with H$\alpha$ measurement. \citet{2012ApJ...755L..29B} 
used [OII] $\lambda3727$ to estimate the SFR of the ultra-luminous PS1-12bam host, 
which at $z=1.566$ has H$\alpha$ shifted to the NIR. The agreement between the SFR 
estimates from [OII] $\lambda3727$ and H$\alpha$ for the host of SN~2010gx supports 
[OII] $\lambda3727$ being reliable for these types of galaxies at higher redshift. We 
detected the galaxy interstellar medium (ISM) absorption of the Mg II $\lambda\lambda2796/2803$ doublet in 
the Gemini spectrum of 1 Apr. 2010 \citep{2010ApJ...724L..16P}. The equivalent widths are 
$1.6\pm0.2$ and $2.6\pm0.3$\,\AA\ respectively, giving a ratio $W_{\rm 2796}/W_{\rm 2803} = 0.6$. 
The strength of the line $\lambda2803$ component is comparable with that seen toward GRB lines of
site and somewhat higher than observed for PS1-12bam \citep{2012ApJ...755L..29B}. The line ratio is
significantly smaller than 2 (the ratio of their oscillator strengths) suggesting we are not on the linear 
regime, nor the square root regime on the curve of growth. Nevertheless the absorption line strengths 
are similar to PS1-12bam and GRB sightlines. The MgII line centroids give identical redshifts 
($z=0.231$) to the [OIII] emission lines ($z=0.230$) detected in the same spectrum. 

The profile of the host galaxy appears slightly broader than the
  stellar point-spread-function on the best seeing images, hence we
  can estimate a physical diameter of the extended source. 
We measured the FWHM both of the host galaxy ($\sim0.73''$) and the average ($\sim0.58''$) of 10 reference stars 
in $r-$band image taken on 14 January 2011. Assuming the relation 
(galaxy observed FWHM)$^{2} \simeq$ (PSF FWHM)$^{2} + $(intrinsic galaxy FWHM)$^{2}$, we estimate a
physical diameter of $\sim 2.4$ kpc  at thus redshift.
The $g-$band absolute magnitude of the host galaxy is $M_{g}=-17.42\pm0.17$, after applying the 
Milky Way extinction of 0.13 \citep{2011ApJ...737..103S}, a k-correction $0.27\pm0.08$ 
\citep{2010MNRAS.405.1409C} and a correction for internal dust extinction of $\sim0.5$ mag. 
The latter comes from the measured intrinsic H$\alpha$/H$\beta$ ratio and assuming $R_{V}=3.16$ 
(applicable for an LMC environment ; \citealt{1992ApJ...395..130P})\footnote{This would be reduced by 
0.05 mag if we used a value suitable for an SMC environment; R$_{V}=2.93$ \citep{1992ApJ...395..130P}.}

The observed photometric $g-r$, $r-i$ colour terms were corrected for Milky Way extinction 
\citep{2011ApJ...737..103S} and used to fit the SED of a stellar population model 
\citep{2005MNRAS.362..799M}. We used the low-metallicity (Z$\sim1/20\,Z_{\odot}$) models, 
assuming a Salpeter initial mass function with red horizontal branch morphology. The measured colours 
give reasonable agreement with a population model of age between 20 to 30 Myr. We further employed 
the {\sc MAGPHYS} galaxy SED models of \citet{2008MNRAS.388.1595D}. Fig.\,1 shows the observed 
SED and the redshifted {\sc MAGPHYS} best-fit ($\chi^{2}=0.58$) model spectrum with total stellar mass 
of $1.6\times10^{8}\,M_{\odot}$; the fitting also gives a likelyhood distribution of parameters, the 
$\pm 1\sigma $ values of the likelyhood distribution are $1.0\times10^{8}$ and $2.2\times10^{8}\,M_{\odot}$ 
separately. Hence a specific star formation rate (sSFR) of ($1.9-4.1)\times10^{-9}$\,year$^{-1}$.

\section{Discussion and conclusion}

It has already been demonstrated that the peak luminosity of super-luminous SNe such as SN~2010gx 
cannot be due to $^{56}$Ni \citep{2010ApJ...724L..16P, 2011Natur.474..487Q, 2011ApJ...743..114C}.
Chomiuk et al. illustrated that applying Arnett's rule \citep{1982ApJ...253..785A}, results in an 
unphysical solution, with the mass of $^{56}$Ni exceeding the total C+O ejecta mass, but see 
\citet{2009Natur.462..624G, 2010A&A...512A..70Y}. Our search 
for a luminous tail phase in SN~2010gx was not motivated by what powers the peak luminosity, rather 
it is to determine if there is any sign of radioactive $^{56}$Co at this stage and any similarity to 
$^{56}$Ni-rich Ic SNe such as SN~1998bw. The only robust conclusion we can draw is that there is 
unlikely to be any excessive mass of $^{56}$Co above and beyond about $0.4\,M_{\odot}$ (Fig.\,2). This 
limit on the late luminosity will also allow restrictions on the magnetar models of 
\citet{2010ApJ...717..245K} when put in context with other super-luminous supernovae of similar types 
(Inserra et al. in prep).

Our detection of the [OIII] $\lambda4363$ auroral line gives confidence to our measurement of the 
remarkably low metallicity of  $12+\log {\rm(O/H)}=7.46\pm0.10$ dex ($0.06\,Z_{\odot}$). 
In Fig.\,3 we show a compilation of other dwarf galaxies with the [OIII] $\lambda4363$ direct method 
measurements, including four GRB hosts from \citet{2008AJ....135.1136M}. The host of SN~2010gx 
is on the lower end and is amongst the lowest metallicity dwarf galaxies known. It has been clear from 
the early discoveries of these SNe at $z<1$ 
\citep{2010ApJ...724L..16P,2011Natur.474..487Q,2011ApJ...743..114C}, that the hosts are faint dwarf 
galaxies and \citet{2011ApJ...727...15N} showed that they have high sSFRs. If the extremely low value 
we determine is a common factor amongst these SNe then it will be a major constraint on the progenitor 
channel. 

The magnetar scenario suggested by \citet{2010ApJ...717..245K} to explain these SNe may benefit from 
low metallicity progenitors. Massive stars rotate more rapidly at SMC metallicity ($0.2\,Z_{\odot}$) 
than solar \citep{2007A&A...462..683M}, although $0.06\,Z_{\odot}$ is unexplored territory. 
Evolutionary models of rotating stars have focused attention on low metallicity environments as mass-loss 
is reduced \citep{2007A&A...473..603M} hence angular momentum loss is lower 
\citep{2000ARA&A..38..143M}. The energy input required from a magnetar requires the neutron star to be 
both highly magnetic ($B\sim10^{14}$\,G) and rotating rapidly at formation ($P_{i}\sim2-20$\,ms). It must 
also occur in a Wolf-Rayet or carbon-oxygen star, since there is no sign of hydrogen in the spectra of 
SN~2010gx or any similar SNe. \citet{2012Sci...337..927G} gives an estimated relative rate of the number 
of 2010gx-like SNe compared to the total core-collapse SN rate of $10^{-4}$. This is probably uncertain 
by a factor of a few, and is more likely to increase due to previous events not being recognised as highly 
luminous. If the progenitor channel requires low metallicity (as we measure), high mass (to produce a WC star) 
and fast rotation (to produce a rapidly rotating magnetar) then this very low rate may not be unexpected.
\citet{2008A&A...489..359Y} estimate that $\sim$1\% of the star formation in the Local Universe ($z<0.04$) 
is in galaxies with $Z<0.2\,Z_{\odot}$; around 13\% of massive stars above the SN threshold of 
$8\,M_{\odot}$ are $>60\,M_{\odot}$ (assuming a Salpeter IMF); and about 10\% of O-stars in the SMC rotate 
faster than $\sim$300\,km\,s$^{-1}$ \citep{2007A&A...473..603M}. One can speculate that all of these three
conditions are required to produce the rate of the 2010gx-like SNe and the estimated rate (which is itself 
uncertain) would be qualitatively reproduced. The high-mass requirement could also be substituted for an 
interacting binary system which causes spin-up in the CO-core that collapses, similar to the proposed GRB 
systems \citep{2007A&A...465L..29C}. The \citet{2011ApJ...729L...6C} scenario of interaction with a dense
wind would probably require extreme progenitor mass and pulsational mass-loss. The fact that we see
Wolf-Rayet stars at very low metallicity (e.g. in IZw13 at $0.02\,Z_{\odot}$) \citep{2002ApJ...579L..75B} 
could suggest that pulsational mass-loss is more prevalent  at low metallicity. This one measurement is
intriguing new information on these extreme SNe but further measurements of the metallicity, star formation 
rate and volumetric rates of these supernovae are required to constrain their explosion channels further. 

\section*{Acknowledgments} 
S.J.S acknowledges support from a  European Research Council Advanced Grant. T.-W. Chen thanks 
Meng-Chun Tsai, Zheng Zheng, Kai-Lung Sun, Christy Tremonti, Elisabete da Cunha, Stephane Charlot 
and Maryam Modjaz for useful advice. {\it Facility:} \facility{Gemini:South (GMOS)}, \facility{Gemini:Gillett (GMOS)}\\


\begin{figure*}
\mbox{
\subfigure{
\includegraphics[width=0.3\textwidth,angle=0]{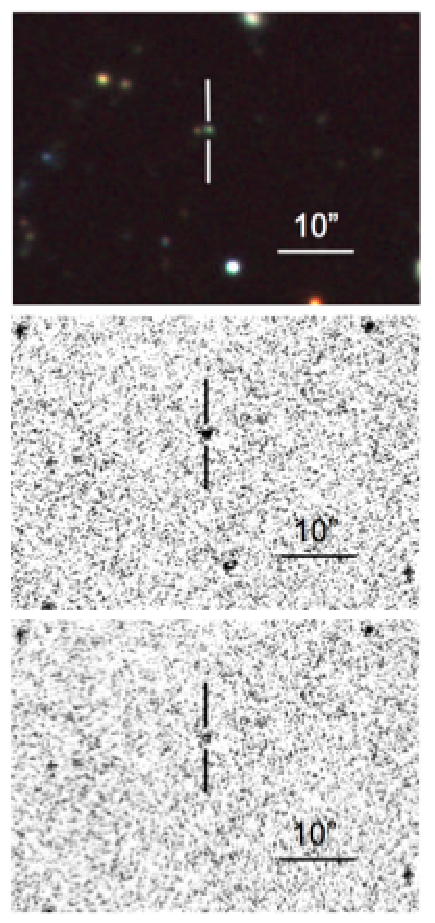}
\label{fig1:subfig1}
}
\subfigure{
\includegraphics[width=0.7\textwidth,angle=0]{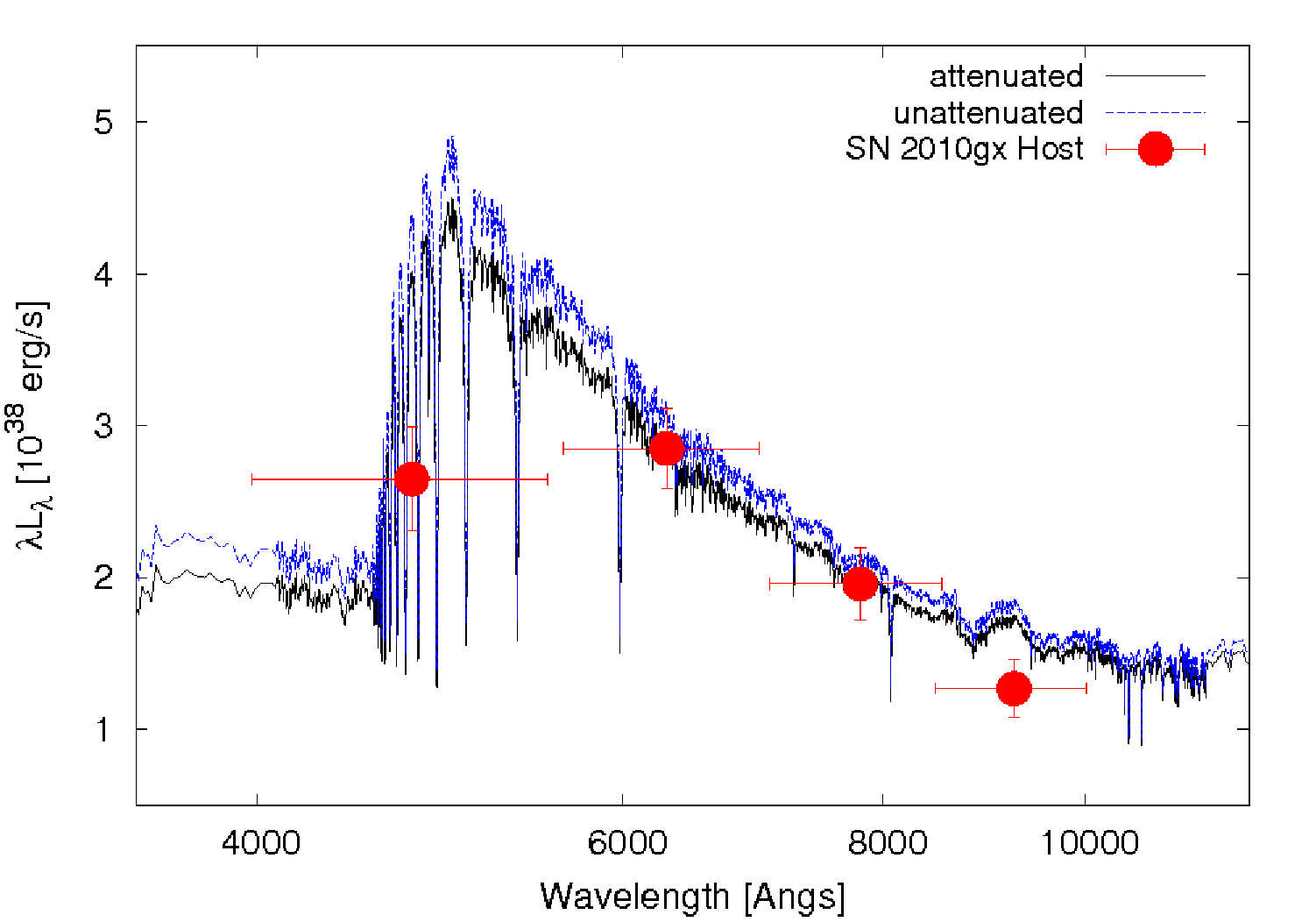}
\label{fig1:subfig2}
}
}
\caption{\textsl{Left} : upper- Colour combined ($gri$) Gemini GMOS images of the host galaxy of 
SN~2010gx; middle- A fake SN 5$\sigma$ detection in r-band after mage subtraction; bottom- 
A fake 3$\sigma$ detection. \textsl{Right}: Photometry of the host galaxy (red circle) with the 
best-fit ($\chi^{2}=0.58$) model galaxy SED (black line) from MAGPHYS \citep{2008MNRAS.388.1595D}, 
the blue line shows the same model without attenuation by dust.}
\label{fig1}
\end {figure*}

\begin{figure*}
\includegraphics[angle=270,width=\linewidth]{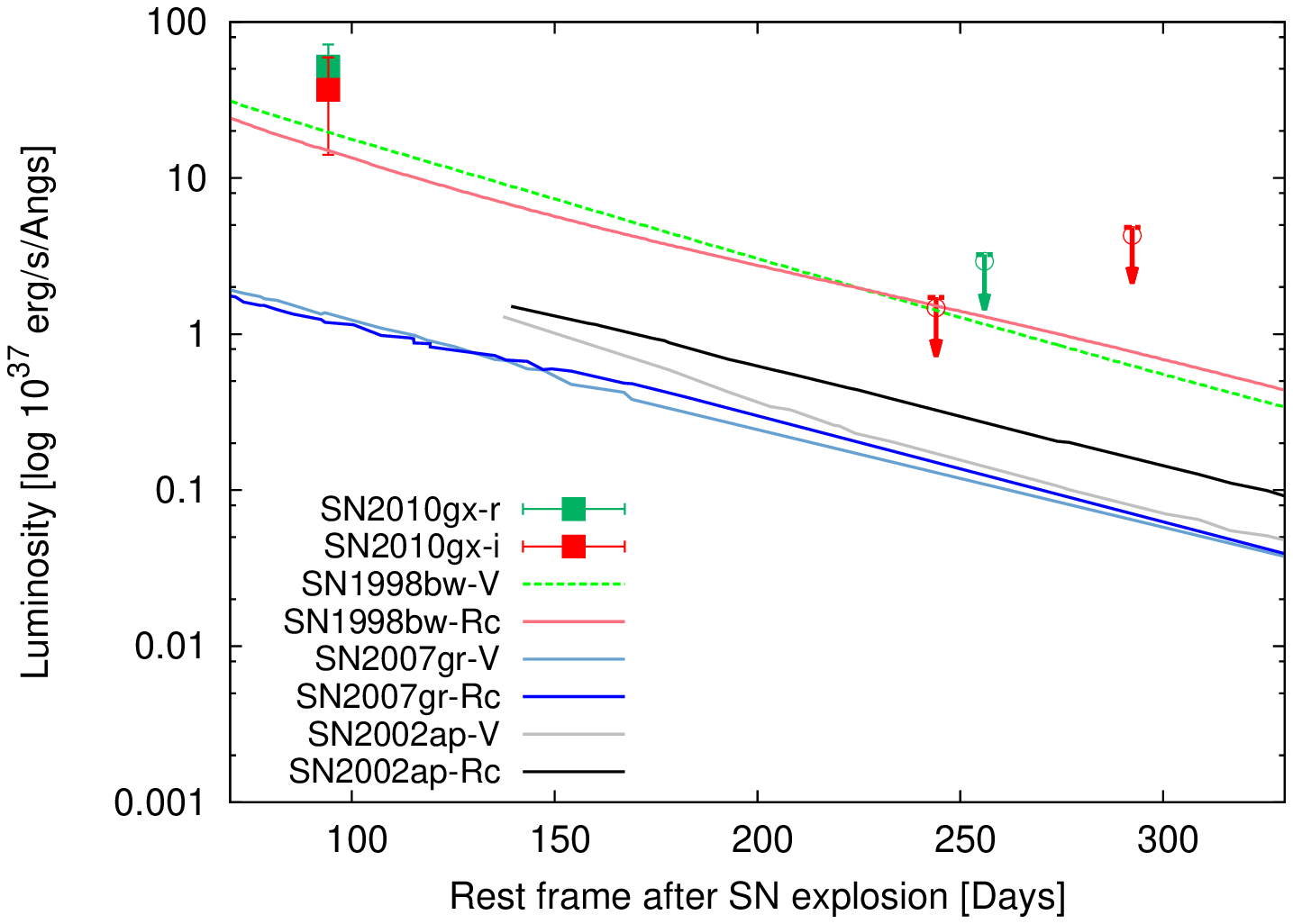}
\caption{The limiting luminosities of SN~2010gx (\citealt{2010ApJ...724L..16P}, this work) 
compared with SN~1998bw\citep{2011AJ....141..163C}, SN~2002ap\citep{2006ApJ...644..400T} 
and SN~2007gr \citep{2009A&A...508..371H}. At 244 days we have a flux limit similar to that of 
SN~1998bw.}
\label{fig2}
\end{figure*}

\begin{figure*}
\center{
\subfigure{
\includegraphics[width=0.5\linewidth,angle=-90]{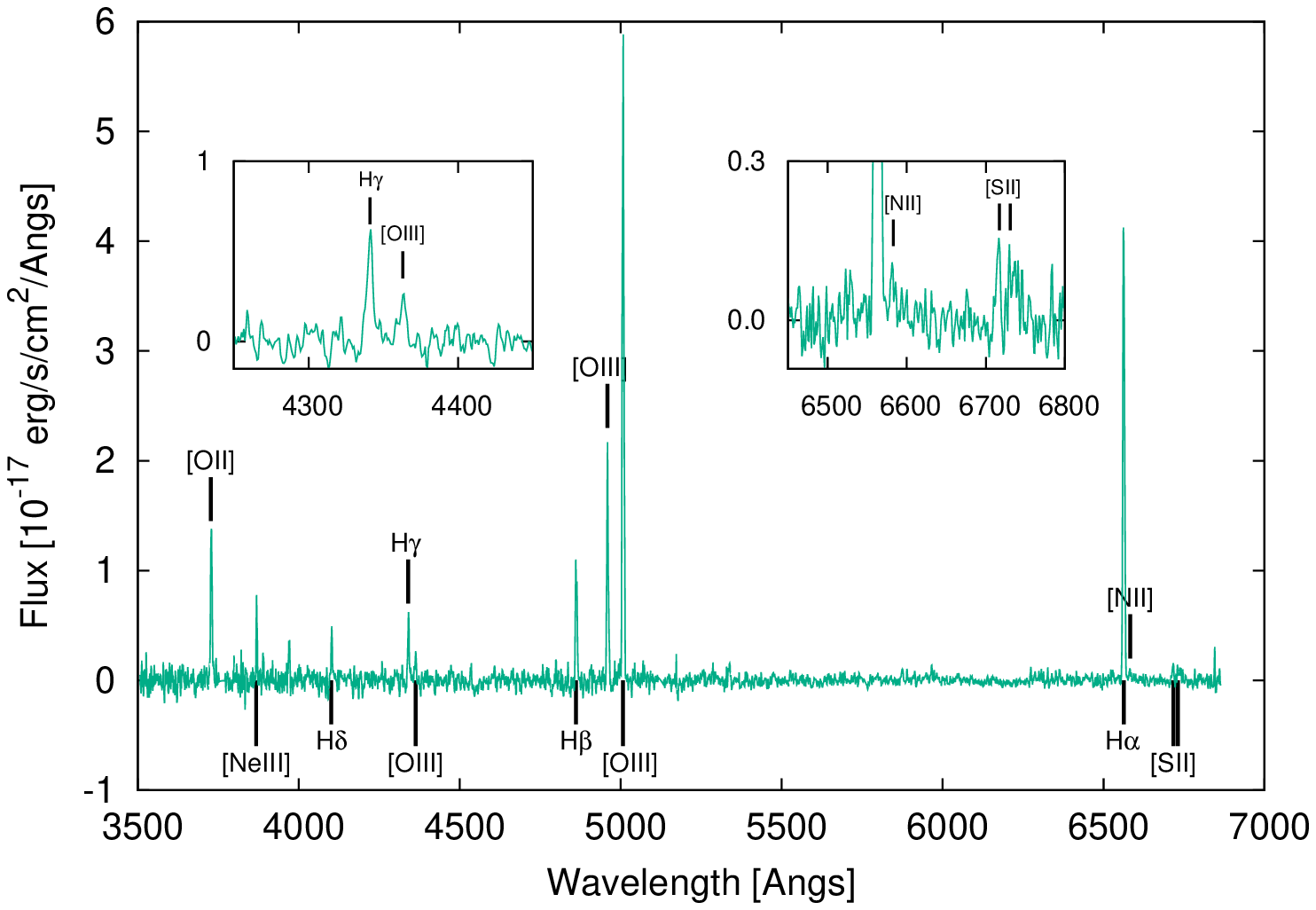}
\label{fig3:subfig1}
}
\subfigure{
\includegraphics[width=0.5\linewidth,angle=-90]{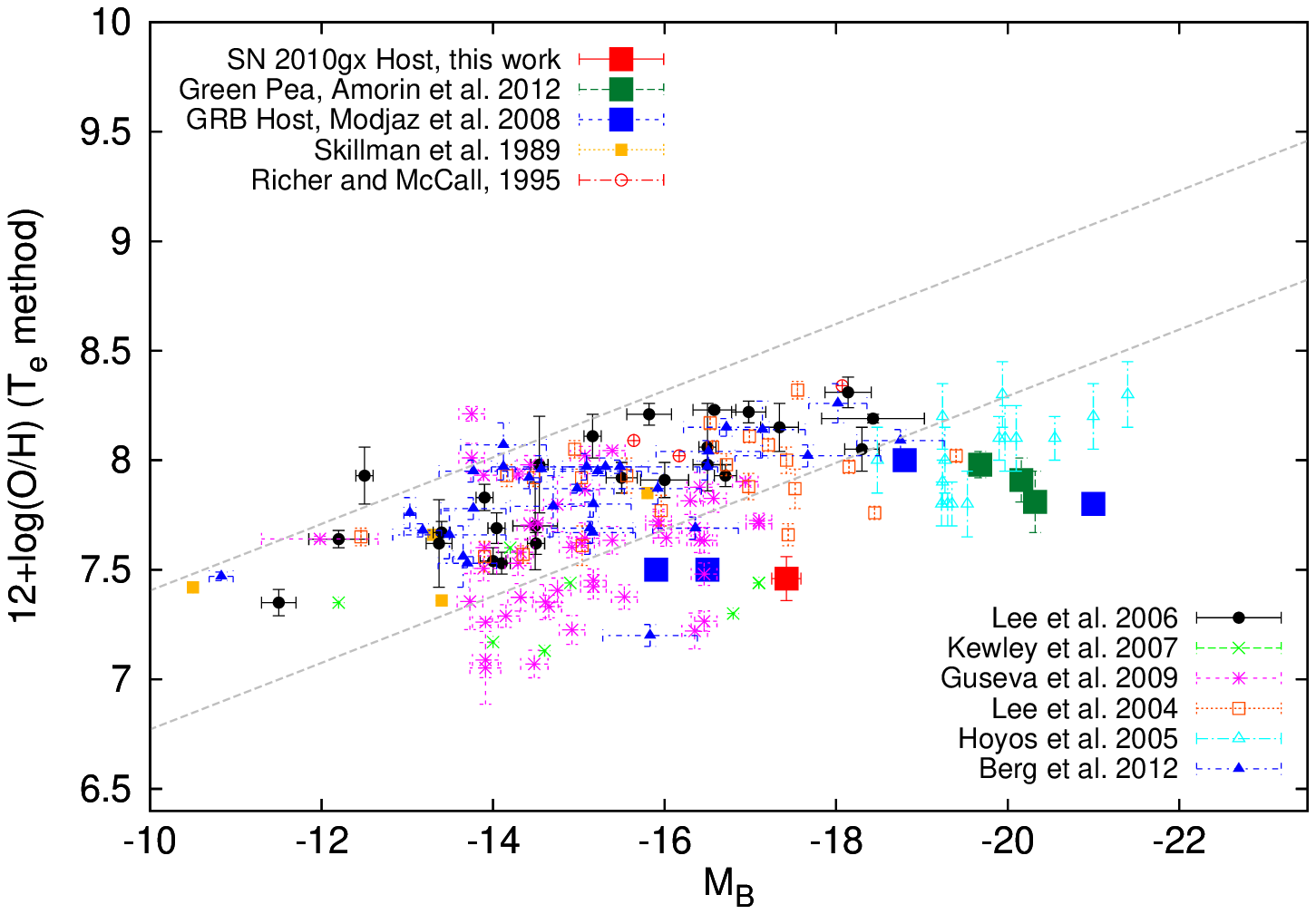}
\label{fig3:subfig2}
}
}
\caption{\textsl{Upper}: Spectrum of the host galaxy of SN~2010gx. The [OIII] $\lambda4363$ 
auroral line is shown in the inset. \textsl{Bottom}: Luminosity-metallicity relationship for dwarf 
galaxies in the local Universe, showing the metallicity as measured using the direct (T$_{e}$) 
method, and using data from the literature listed in the figure key \citep{1989ApJ...347..875S, 
1995ApJ...445..642R, 2004ApJ...616..752L, 2005ApJ...635L..21H, 2006ApJ...647..970L, 
2007AJ....133..882K, 2008AJ....135.1136M, 2009A&A...505...63G, 2012ApJ...749..185A, 
2012ApJ...754...98B}. Grey lines are the boundary 
of normal galaxy mass-metallicity relation adapted from \citet{2012ApJ...747...15K}, with 
metallicities calculated using blue supergiants. }
\label{fig3}
\end {figure*}


\begin{thebibliography}{50}
\expandafter\ifx\csname natexlab\endcsname\relax\def\natexlab#1{#1}\fi

\bibitem[{{Amor{\'{\i}}n} {et~al.}(2012){Amor{\'{\i}}n}, {P{\'e}rez-Montero},
  {V{\'{\i}}lchez}, \& {Papaderos}}]{2012ApJ...749..185A}
{Amor{\'{\i}}n}, R., {P{\'e}rez-Montero}, E., {V{\'{\i}}lchez}, J.~M., \&
  {Papaderos}, P. 2012, \apj, 749, 185

\bibitem[{{Arnett}(1982)}]{1982ApJ...253..785A}
{Arnett}, W.~D. 1982, \apj, 253, 785

\bibitem[{{Asplund} {et~al.}(2009){Asplund}, {Grevesse}, {Sauval}, \&
  {Scott}}]{2009ARA&A..47..481A}
{Asplund}, M., {Grevesse}, N., {Sauval}, A.~J., \& {Scott}, P. 2009, \araa, 47,
  481

\bibitem[{{Berg} {et~al.}(2012){Berg}, {Skillman}, {Marble}, {van Zee},
  {Engelbracht}, {Lee}, {Kennicutt}, {Calzetti}, {Dale}, \&
  {Johnson}}]{2012ApJ...754...98B}
{Berg}, D.~A., {Skillman}, E.~D., {Marble}, A.~R., {et~al.} 2012, \apj, 754, 98

\bibitem[{{Berger} {et~al.}(2012){Berger}, {Chornock}, {Lunnan}, {Foley},
  {Czekala}, {Rest}, {Leibler}, {Soderberg}, {Roth}, {Narayan}, {Huber},
  {Milisavljevic}, {Sanders}, {Drout}, {Margutti}, {Kirshner}, {Marion},
  {Challis}, {Riess}, {Smartt}, {Burgett}, {Hodapp}, {Heasley}, {Kaiser},
  {Kudritzki}, {Magnier}, {McCrum}, {Price}, {Smith}, {Tonry}, \&
  {Wainscoat}}]{2012ApJ...755L..29B}
{Berger}, E., {Chornock}, R., {Lunnan}, R., {et~al.} 2012, \apjl, 755, L29

\bibitem[{{Blinnikov} \& {Sorokina}(2010)}]{2010arXiv1009.4353B}
{Blinnikov}, S.~I., \& {Sorokina}, E.~I. 2010, ArXiv: 1009.4353

\bibitem[{{Bresolin} {et~al.}(2009){Bresolin}, {Gieren}, {Kudritzki},
  {Pietrzy{\'n}ski}, {Urbaneja}, \& {Carraro}}]{2009ApJ...700..309B}
{Bresolin}, F., {Gieren}, W., {Kudritzki}, R.-P., {et~al.} 2009, \apj, 700, 309

\bibitem[{{Brown} {et~al.}(2002){Brown}, {Heap}, {Hubeny}, {Lanz}, \&
  {Lindler}}]{2002ApJ...579L..75B}
{Brown}, T.~M., {Heap}, S.~R., {Hubeny}, I., {Lanz}, T., \& {Lindler}, D. 2002,
  \apjl, 579, L75

\bibitem[{{Cantiello} {et~al.}(2007){Cantiello}, {Yoon}, {Langer}, \&
  {Livio}}]{2007A&A...465L..29C}
{Cantiello}, M., {Yoon}, S.-C., {Langer}, N., \& {Livio}, M. 2007, \aap, 465,
  L29

\bibitem[{{Chevalier} \& {Irwin}(2011)}]{2011ApJ...729L...6C}
{Chevalier}, R.~A., \& {Irwin}, C.~M. 2011, \apjl, 729, L6

\bibitem[{{Chilingarian} {et~al.}(2010){Chilingarian}, {Melchior}, \&
  {Zolotukhin}}]{2010MNRAS.405.1409C}
{Chilingarian}, I.~V., {Melchior}, A.-L., \& {Zolotukhin}, I.~Y. 2010, \mnras,
  405, 1409

\bibitem[{{Chomiuk} {et~al.}(2011){Chomiuk}, {Chornock}, {Soderberg}, {Berger},
  {Chevalier}, {Foley}, {Huber}, {Narayan}, {Rest}, {Gezari}, {Kirshner},
  {Riess}, {Rodney}, {Smartt}, {Stubbs}, {Tonry}, {Wood-Vasey}, {Burgett},
  {Chambers}, {Czekala}, {Flewelling}, {Forster}, {Kaiser}, {Kudritzki},
  {Magnier}, {Martin}, {Morgan}, {Neill}, {Price}, {Roth}, {Sanders}, \&
  {Wainscoat}}]{2011ApJ...743..114C}
{Chomiuk}, L., {Chornock}, R., {Soderberg}, A.~M., {et~al.} 2011, \apj, 743,
  114

\bibitem[{{Clocchiatti} {et~al.}(2011){Clocchiatti}, {Suntzeff}, {Covarrubias},
  \& {Candia}}]{2011AJ....141..163C}
{Clocchiatti}, A., {Suntzeff}, N.~B., {Covarrubias}, R., \& {Candia}, P. 2011,
  \aj, 141, 163

\bibitem[{{da Cunha} {et~al.}(2008){da Cunha}, {Charlot}, \&
  {Elbaz}}]{2008MNRAS.388.1595D}
{da Cunha}, E., {Charlot}, S., \& {Elbaz}, D. 2008, \mnras, 388, 1595

\bibitem[{{Gal-Yam}(2012)}]{2012Sci...337..927G}
{Gal-Yam}, A. 2012, Science, 337, 927

\bibitem[{{Gal-Yam} {et~al.}(2009){Gal-Yam}, {Mazzali}, {Ofek}, {Nugent},
  {Kulkarni}, {Kasliwal}, {Quimby}, {Filippenko}, {Cenko}, {Chornock},
  {Waldman}, {Kasen}, {Sullivan}, {Beshore}, {Drake}, {Thomas}, {Bloom},
  {Poznanski}, {Miller}, {Foley}, {Silverman}, {Arcavi}, {Ellis}, \&
  {Deng}}]{2009Natur.462..624G}
{Gal-Yam}, A., {Mazzali}, P., {Ofek}, E.~O., {et~al.} 2009, \nat, 462, 624

\bibitem[{{Guseva} {et~al.}(2009){Guseva}, {Papaderos}, {Meyer}, {Izotov}, \&
  {Fricke}}]{2009A&A...505...63G}
{Guseva}, N.~G., {Papaderos}, P., {Meyer}, H.~T., {Izotov}, Y.~I., \& {Fricke},
  K.~J. 2009, \aap, 505, 63

\bibitem[{{Hoyos} {et~al.}(2005){Hoyos}, {Koo}, {Phillips}, {Willmer}, \&
  {Guhathakurta}}]{2005ApJ...635L..21H}
{Hoyos}, C., {Koo}, D.~C., {Phillips}, A.~C., {Willmer}, C.~N.~A., \&
  {Guhathakurta}, P. 2005, \apjl, 635, L21

\bibitem[{{Hunter} {et~al.}(2009){Hunter}, {Valenti}, {Kotak}, {Meikle},
 {Taubenberger}, {Pastorello}, {Benetti}, {Stanishev}, {Smartt}, {Trundle}, 
 {Arkharov}, {Bufano}, {Cappellaro}, {Di Carlo}, {Dolci}, {Elias-Rosa},
 {Frandsen}, {Fynbo}, {Hopp}, {Larionov}, {Laursen}, {Mazzali}, {Navasardyan},
 {Ries}, {Riffeser}, {Rizzi}, {Tsvetkov}, {Turatto}, \& {Wilke}}]{2009A&A...508..371H}
{Hunter}, D.~J., {Valenti}, S., {Kotak}, R., {et~al.}. 2009, \aap, 508, 371

\bibitem[{{Kaiser} {et~al.}(2010){Kaiser}, {Burgett}, {Chambers}, {Denneau},
  {Heasley}, {Jedicke}, {Magnier}, {Morgan}, {Onaka}, \&
  {Tonry}}]{2010SPIE.7733E..12K}
{Kaiser}, N., {Burgett}, W., {Chambers}, K., {et~al.}. 2010, SPIE, 7733, 12
  
\bibitem[{{Kasen} \& {Bildsten}(2010)}]{2010ApJ...717..245K}
{Kasen}, D., \& {Bildsten}, L. 2010, \apj, 717, 245

\bibitem[{{Kennicutt}(1998)}]{1998ARA&A..36..189K}
{Kennicutt}, Jr., R.~C. 1998, \araa, 36, 189

\bibitem[{{Kewley} {et~al.}(2007){Kewley}, {Brown}, {Geller}, {Kenyon}, \&
  {Kurtz}}]{2007AJ....133..882K}
{Kewley}, L.~J., {Brown}, W.~R., {Geller}, M.~J., {Kenyon}, S.~J., \& {Kurtz},
  M.~J. 2007, \aj, 133, 882

\bibitem[{{Kudritzki} {et~al.}(2012){Kudritzki}, {Urbaneja}, {Gazak},
  {Bresolin}, {Przybilla}, {Gieren}, \&
  {Pietrzy{\'n}ski}}]{2012ApJ...747...15K}
{Kudritzki}, R.-P., {Urbaneja}, M.~A., {Gazak}, Z., {et~al.} 2012, \apj, 747,
  15

\bibitem[{{Lee} {et~al.}(2006){Lee}, {Skillman}, {Cannon}, {Jackson}, {Gehrz},
  {Polomski}, \& {Woodward}}]{2006ApJ...647..970L}
{Lee}, H., {Skillman}, E.~D., {Cannon}, J.~M., {et~al.} 2006, \apj, 647, 970

\bibitem[{{Lee} {et~al.}(2004){Lee}, {Salzer}, \&
  {Melbourne}}]{2004ApJ...616..752L}
{Lee}, J.~C., {Salzer}, J.~J., \& {Melbourne}, J. 2004, \apj, 616, 752

\bibitem[{{Maeda} {et~al.}(2006){Maeda}, {Nomoto}, {Mazzali}, \&
  {Deng}}]{2006ApJ...640..854M}
{Maeda}, K., {Nomoto}, K., {Mazzali}, P.~A., \& {Deng}, J. 2006, \apj, 640, 854

\bibitem[{{Maeder} \& {Meynet}(2000)}]{2000ARA&A..38..143M}
{Maeder}, A., \& {Meynet}, G. 2000, \araa, 38, 143

\bibitem[{{Maraston}(2005)}]{2005MNRAS.362..799M}
{Maraston}, C. 2005, \mnras, 362, 799

\bibitem[{{Martayan} {et~al.}(2007){Martayan}, {Fr{\'e}mat}, {Hubert},
  {Floquet}, {Zorec}, \& {Neiner}}]{2007A&A...462..683M}
{Martayan}, C., {Fr{\'e}mat}, Y., {Hubert}, A.-M., {et~al.} 2007, \aap, 462,
  683

\bibitem[{{Modjaz} {et~al.}(2008){Modjaz}, {Kewley}, {Kirshner}, {Stanek},
  {Challis}, {Garnavich}, {Greene}, {Kelly}, \& {Prieto}}]{2008AJ....135.1136M}
{Modjaz}, M., {Kewley}, L., {Kirshner}, R.~P., {et~al.} 2008, \aj, 135, 1136

\bibitem[{{Mokiem} {et~al.}(2007){Mokiem}, {de Koter}, {Vink}, {Puls}, {Evans},
  {Smartt}, {Crowther}, {Herrero}, {Langer}, {Lennon}, {Najarro}, \&
  {Villamariz}}]{2007A&A...473..603M}
{Mokiem}, M.~R., {de Koter}, A., {Vink}, J.~S., {et~al.} 2007, \aap, 473, 603

\bibitem[{{Neill} {et~al.}(2011){Neill}, {Sullivan}, {Gal-Yam}, {Quimby},
  {Ofek}, {Wyder}, {Howell}, {Nugent}, {Seibert}, {Martin}, {Overzier},
  {Barlow}, {Foster}, {Friedman}, {Morrissey}, {Neff}, {Schiminovich},
  {Bianchi}, {Donas}, {Heckman}, {Lee}, {Madore}, {Milliard}, {Rich}, \&
  {Szalay}}]{2011ApJ...727...15N}
{Neill}, J.~D., {Sullivan}, M., {Gal-Yam}, A., {et~al.} 2011, \apj, 727, 15

\bibitem[{{Osterbrock}(1989)}]{1989agna.book.....O}
{Osterbrock}, D.~E. 1989, {Astrophysics of gaseous nebulae and active galactic
  nuclei}

\bibitem[{{Pastorello} {et~al.}(2010){Pastorello}, {Smartt}, {Botticella},
  {Maguire}, {Fraser}, {Smith}, {Kotak}, {Magill}, {Valenti}, {Young},
  {Gezari}, {Bresolin}, {Kudritzki}, {Howell}, {Rest}, {Metcalfe}, {Mattila},
  {Kankare}, {Huang}, {Urata}, {Burgett}, {Chambers}, {Dombeck}, {Flewelling},
  {Grav}, {Heasley}, {Hodapp}, {Kaiser}, {Luppino}, {Lupton}, {Magnier},
  {Monet}, {Morgan}, {Onaka}, {Price}, {Rhoads}, {Siegmund}, {Stubbs},
  {Sweeney}, {Tonry}, {Wainscoat}, {Waterson}, {Waters}, \&
  {Wynn-Williams}}]{2010ApJ...724L..16P}
{Pastorello}, A., {Smartt}, S.~J., {Botticella}, M.~T., {et~al.} 2010, \apjl,
  724, L16

\bibitem[{{Pei}(1992)}]{1992ApJ...395..130P}
{Pei}, Y.~C. 1992, \apj, 395, 130

\bibitem[{{P{\'e}rez-Montero} \& {D{\'{\i}}az}(2003)}]{2003MNRAS.346..105P}
{P{\'e}rez-Montero}, E., \& {D{\'{\i}}az}, A.~I. 2003, \mnras, 346, 105

\bibitem[{{Quimby} {et~al.}(2011){Quimby}, {Kulkarni}, {Kasliwal}, {Gal-Yam},
  {Arcavi}, {Sullivan}, {Nugent}, {Thomas}, {Howell}, {Nakar}, {Bildsten},
  {Theissen}, {Law}, {Dekany}, {Rahmer}, {Hale}, {Smith}, {Ofek}, {Zolkower},
  {Velur}, {Walters}, {Henning}, {Bui}, {McKenna}, {Poznanski}, {Cenko}, \&
  {Levitan}}]{2011Natur.474..487Q}
{Quimby}, R.~M., {Kulkarni}, S.~R., {Kasliwal}, M.~M., {et~al.} 2011, \nat,
  474, 487

\bibitem[{{Rau} {et~al.}(2009){Rau}, {Kulkarni}, {Law}, {Bloom}, {Ciardi},
  {Djorgovski}, {Fox}, {Gal-Yam}, {Grillmair}, {Kasliwal}, {Nugent}, {Ofek},
  {Quimby}, {Reach}, {Shara}, {Bildsten}, {Cenko}, {Drake}, {Filippenko},
  {Helfand}, {Helou}, {Howell}, {Poznanski}, \&
  {Sullivan}}]{2009PASP..121.1334R}
{Rau}, A., {Kulkarni}, S.~R., {Law}, N.~M., {et~al.} 2009, \pasp, 121, 1334

\bibitem[{{Richer} \& {McCall}(1995)}]{1995ApJ...445..642R}
{Richer}, M.~G., \& {McCall}, M.~L. 1995, \apj, 445, 642

\bibitem[{{Schlafly} \& {Finkbeiner}(2011)}]{2011ApJ...737..103S}
{Schlafly}, E.~F., \& {Finkbeiner}, D.~P. 2011, \apj, 737, 103


\bibitem[{{Skillman} {et~al.}(1989){Skillman}, {Kennicutt}, \&
  {Hodge}}]{1989ApJ...347..875S}
{Skillman}, E.~D., {Kennicutt}, R.~C., \& {Hodge}, P.~W. 1989, \apj, 347, 875

\bibitem[{{Stoll} {et~al.}(2011){Stoll}, {Prieto}, {Stanek}, {Pogge}, {Szczygie{\l}}, 
{Pojma{\'n}ski}, {Antognini}, \& {Yan}}]{2011ApJ...730...34S} 
{Stoll}, R., {Prieto}, J.~L., {Stanek}, K.~Z. {et~al.} 2011, \apj, 730, 34 

\bibitem[{{Tomita} {et~al.}(2006){Tomita}, {Deng}, {Maeda}, {Yoshii},
{Nomoto}, {Mazzali}, {Suzuki}, {Kobayashi}, {Minezaki}, {Aoki},
{Enya}, \& {Suganuma}}] {2006ApJ...644..400T} {Tomita}, H., {Deng}, J., {Maeda}, K. {et~al.} 2006, \apj, 644, 400

\bibitem[{{Wright}(2006)}]{2006PASP..118.1711W}
{Wright}, E.~L. 2006, \pasp, 118, 1711


\bibitem[{{Young} {et~al.}(2008){Young}, {Smartt}, {Mattila}, {Tanvir},
  {Bersier}, {Chambers}, {Kaiser}, \& {Tonry}}]{2008A&A...489..359Y}
{Young}, D.~R., {Smartt}, S.~J., {Mattila}, S., {et~al.} 2008, \aap, 489, 359

\bibitem[{{Young} {et~al.}(2010){Young}, {Smartt}, {Valenti}, {Pastorello},
  {Benetti}, {Benn}, {Bersier}, {Botticella}, {Corradi}, {Harutyunyan},
  {Hrudkova}, {Hunter}, {Mattila}, {de Mooij}, {Navasardyan}, {Snellen},
  {Tanvir}, \& {Zampieri}}]{2010A&A...512A..70Y}
{Young}, D.~R., {Smartt}, S.~J., {Valenti}, S., {et~al.} 2010, \aap, 512, A70

\end{thebibliography}

\clearpage

\begin{table}
\begin{tabular}[t]{lllllllll}
\hline
\hline
Date & MJD & Phase & g & r & i & z & Telescope & Instrument \\
\hline
2010 Jun 4  & 55351.97 & 74.4 &            &             & 21.13(0.19) & &  GS & GMOS-S \\
2010 Jun 8  & 55354.93 & 76.8 & 23.55(0.28) & 21.52(0.09) & 21.31(0.12) & &  WHT & ACAM  \\
2010 Jun 8  & 55355.48 & 77.2 &            & 21.58(0.13) & 21.31(0.16) &  &  FTS  & EM03 \\
2010 Jun 13 & 55360.47 & 81.3 & 23.91(0.30) & 21.81(0.20) & 21.56(0.13) & &  FTS & EM03 \\
2010 Jun 13 & 55360.92 & 81.6 & 23.95(0.49) & 21.83(0.12) & 21.58(0.29) & & LT & RATcam \\
2010 Jun 16 & 55363.48 & 83.7 & $>$23.73    & 21.91(0.16) &             & &  FTS & EM03 \\
2010 Jun 29 & 55376.38 & 94.2 &            & 22.47(0.21) & 22.38(0.33) & &  FTS & EM03 \\
2010 Dec 30 & 55560.63 & 244.0 &            &             & H 22.90(0.13) & H 22.98(0.16) &  GN & GMOS-N \\
                    &                 &  &            &            & $>$25.94(0.17)  & $>$24.58(0.29) & &\\ 
2011 Jan 14 & 55575.32 & 256.0 &  H 23.71(0.14) & H 22.98(0.10) &             & & GS  & GMOS-S\\
                    &                 & & $>$26.42(0.18)&  $>$25.68(0.11) &  &  & & \\ 
2011 Feb 28 & 55620.12 & 292.4 &            &             & H 23.20(0.13) & H 23.00(0.13) &  GS & GMOS-S \\
 &  &             &             & & $>$24.78(0.13) & $>$23.35(0.32) &  & \\
2011 May 20 & 55701.28 & 358.4 & H 23.72(0.13) &             &             & &  GN & GMOS-N \\
 &  & & $>$26.00(0.15) &      &                 & &  & \\
2011 May 25 & 55706.08 & 362.3 &          & H 23.08(0.08) & H 23.02(0.11) & H 23.06(0.16) &  GS & GMOS-S \\ 
 &  &             & & $>$25.35(0.15) & $>$24.60(0.20) & &  & \\
2012 Jan 18 & 55944.22 & 555.7 & & H 23.03(0.10) & & &  GS & GMOS-S \\
2012 Jan 20 & 55946.23 & 557.5 & & & H 23.17(0.11) & &  GS & GMOS-S \\
2012 Jan 21 & 55947.21 & 558.3 & H 23.62(0.11) & & & &  GS & GMOS-S \\
\hline
\end{tabular}
\caption{Observed photometry of SN 2010gx and its host galaxy. Data
  until 2010 Jun 29 are from \citet{2010ApJ...724L..16P}. The symbol "H"
  indicates host galaxy photometry, and the "$>$" expresses the
  limiting magnitude of the SN (5$\sigma$). Phase has been corrected for time 
dilation, based on the SN explosion at MJD 55260.5.}
\end{table}

\begin{table}
\begin{tabular}[t]{lllllllllllll}
\hline
\hline
SDSS name & SDSS J112546.72-084942.0 \\
RA & 11:25:46.72 \\
Dec & -08:49:42.0 \\
Redshift & 0.23 \\
Apparent $g$ (mag) & 23.71 $\pm$ 0.14 \\ 	
Apparent $r$ (mag) & 22.98 $\pm$ 0.10 \\
Apparent $i$ (mag) & 22.90 $\pm$ 0.13 \\
Apparent $z$ (mag) & 22.98 $\pm$ 0.16 \\
Galactic extinction$_{g}$ (mag) & 0.13 \\
k-correction$_{g}$ (mag) & 0.27 $\pm$ 0.08 \\
Internal extinction$_{g}$ (mag) & $\sim$ 0.5 (R$_{V}$ = 3.16)\\
Luminosity distance (Mpc) & 1114.7 \\
Absolute $g$ (mag) & -17.42 $\pm$ 0.17 \\
Physical Diameter (kpc) & 2.4 \\ 
Luminosity (H$\alpha$) (erg s$^{-1}$) & $5.20\times10^{40}$ \\
SFR ($M_{\odot}$ year$^{-1}$) & 0.41 \\
Stellar mass ($M_{\odot}$) & $(1.0-2.2)\times10^{8}$ \\
sSFR (year$^{-1}$) & $(1.9-4.1)\times10^{-9}$ \\
$12+\log {\rm(O/H)}$ (T$_{e}$ method) & 7.46 $\pm$ 0.1 \\
\hline 
\hline
Line & Observed Flux $\pm$ Error \\
 & (erg s$^{-1}$ cm$^{-2}$) \\
\hline
$[$OII$]$ $\lambda$3727 & $(1.30\pm0.04)\times10^{-16}$\\
$[$NeIII$]$ $\lambda$3868 & $(6.12\pm0.27)\times10^{-17}$\\
H$\delta$ $\lambda$4102 & $(4.25\pm0.30)\times10^{-17}$ \\
H$\gamma$ $\lambda$4340 & $(5.27\pm0.36)\times10^{-17}$ \\
$[$OIII$]$ $\lambda$4363 &  $(2.27\pm0.23)\times10^{-17}$ \\
H$\beta$ $\lambda$4861 & $(1.29\pm0.03)\times10^{-16}$ \\
$[$OIII$]$ $\lambda$4959 & $(2.13\pm0.04)\times10^{-16}$ \\
$[$OIII$]$ $\lambda$5007 & $(6.57\pm0.08)\times10^{-16}$ \\
H$\alpha$ $\lambda$6563 & $(5.00\pm0.05)\times10^{-16}$ \\
$[$NII$]$ $\lambda$6584 & $(9.81\pm0.82)\times10^{-18}$ \\
$[$SII$]$ $\lambda$6717 & $(1.92\pm0.08)\times10^{-17}$ \\
$[$SII$]$ $\lambda$6731 & $(1.82\pm0.09)\times10^{-17}$ \\
\hline
c(H$\beta$) & 0.48 $\pm 0.04$  \\
-W(H$\beta$) ($\AA$) & 52  \\
\hline
\end{tabular}
\caption {Main properties of the host galaxy of SN~2010gx.}
\end{table}

\end{document}